\documentclass[%
showpacs,
nofootinbib,
 amsmath,amssymb,
 aps,
 twocolumn,
 prl,
 reprint,
floatfix,
]{revtex4-1}

\usepackage{graphicx}
\usepackage{dcolumn}
\usepackage{bm}
\usepackage{amsmath}
\usepackage{amsfonts}
\usepackage{amssymb}
\usepackage{color}
\usepackage{acronym}
\usepackage{multirow}
\usepackage{tabularx}
\usepackage{hyperref}
\usepackage{mathtools}
\usepackage{diagbox}

\hypersetup{
  colorlinks=true,        
  linkcolor=black,         
  citecolor=cyan,         
}

\begin{document}

\preprint{APS/123-QED}

\title{Bayesian parameter estimation using conditional variational autoencoders
for gravitational-wave astronomy}

\author{Hunter Gabbard$^1$}
 \email{Corresponding author: h.gabbard.1@research.gla.ac.uk}
\author{Chris Messenger$^1$}
\author{Ik Siong Heng$^1$}
\author{Francesco Tonolini$^2$}
\author{Roderick Murray-Smith$^2$}

\affiliation{
 SUPA, School of Physics and Astronomy$^1$, \\
 University of Glasgow, \\
 Glasgow G12 8QQ, United Kingdom \\ \\
 School of Computing Science$^2$, \\
 University of Glasgow, \\
 Glasgow G12 8QQ, United Kingdom \\
}

\date{\today}

\maketitle

%

\acrodef{GW}[GW]{Gravitational wave}
\acrodef{BBH}[BBH]{binary black hole}
\acrodef{EM}[EM]{electromagnetic}
\acrodef{CBC}[CBC]{compact binary coalescence}
\acrodef{BNS}[BNS]{binary neutron star}
\acrodef{NSBH}[NSBH]{neutron star black hole}
\acrodef{PSD}[PSD]{power spectral density}
\acrodef{ELBO}[ELBO]{evidence lower bound}
\acrodef{LIGO}[LIGO]{advanced Laser Interferometer Gravitational wave Observatory}
\acrodef{CVAE}[CVAE]{conditional variational autoencoder}
\acrodef{KL}[KL]{Kullback--Leibler}
\acrodef{JS}[JS]{Jensen–-Shannon}
\acrodef{GPU}[GPU]{graphics processing unit}
\acrodef{LVC}[LVC]{LIGO-Virgo Collaboration}
\acrodef{PP}[p-p]{probability-probability}
\acrodef{SNR}[SNR]{signal-to-noise ratio}

%
%
%
\textbf{ 
%
%
\ac{GW} detection is now
commonplace~\cite{PhysRevX.9.031040, PhysRevX.11.021053,PhysRevLett.119.161101} 
and as the
sensitivity of the global network of \ac{GW} detectors improves, we will
observe on the order of 100s of transient \ac{GW} events per
year~\cite{2018LRR....21....3A}. The current methods used to estimate their
source parameters employ optimally sensitive~\cite{2009CQGra..26o5017S} but
computationally costly Bayesian inference approaches~\cite{1409.7215} where
typical analyses have taken between 6 hours and 6 days~\cite{gracedb_O3}.
%
%
For \ac{BNS} and \ac{NSBH} systems prompt counterpart \ac{EM} signatures are
expected on timescales of 1 second -- 1 minute and the current fastest method
for alerting \ac{EM} follow-up observers~\cite{2016PhRvD..93b4013S}, can
provide estimates in on the order of 1 minute, on a limited range of key source
parameters. 
%
%
Here we show that a \ac{CVAE}~\cite{1904.06264,1812.04405} pre-trained on
\ac{BBH} signals can return Bayesian posterior probability estimates. The
training procedure need only be performed once for a given prior parameter
space and the resulting trained machine can then generate samples describing
the posterior distribution $\sim 6$ orders of magnitude faster than existing
techniques.}
%

%
%
%

%
%
The problem of detecting \acp{GW} has largely been solved through the use of
template based matched-filtering, a process recently replicated using machine
learning techniques~\cite{GEORGE201864,PhysRevLett.120.141103,GebKilParHarSch}.
Once a \ac{GW} has been identified through this process, Bayesian inference,
known to be the optimal approach~\cite{2009CQGra..26o5017S}, is used to extract
information about the source parameters of the detected \ac{GW} signal.

%
%
In the standard Bayesian \ac{GW} inference approach, we assume a signal and
noise model and both may have unknown parameters that we are either interested
in inferring or prefer to marginalise away. Each parameter is given a prior
astrophysically motivated probability distribution and in the \ac{GW} case, we
typically assume a Gaussian additive noise model (in reality, the data is not
truly Gaussian). Given a noisy \ac{GW} waveform, we would like to find an
optimal procedure for inferring some set of the unknown \ac{GW} parameters.
Such a procedure should be able to give us an accurate estimate of the
parameters of our observed signal, whilst accounting for the uncertainty
arising from the noise in the data.

%
%
According to Bayes' Theorem, a posterior probability distribution on a set of
parameters, conditional on the measured data, can be represented as
\begin{align}\label{eq:bayes_theorem} 
p(x|y) &\propto p(y|x) p(x), 
\end{align}
where $x$ are the parameters, $y$ is the observed data, $p(x|y)$ is the
posterior, $p(y|x)$ is the likelihood, and $p(x)$ is the prior on the
parameters. The constant of proportionality, which we omit here, is
$p(y)$, the probability of our data, known as the Bayesian evidence or the
marginal likelihood. We typically ignore $p(y)$ since it is a constant and for
parameter estimation purposes we are only interested in the shape of the
posterior.

%
%
Due to the size of the parameter space typically encountered in \ac{GW}
parameter estimation and the volume of data analysed, we must stochastically
sample the parameter space in order to estimate the posterior.  Sampling is
done using a variety of techniques including Nested
Sampling~\cite{skilling2006,cpnest,dynesty} and Markov chain Monte Carlo
methods~\cite{emcee,ptemcee}. The primary software tools used by the \ac{LIGO}
parameter estimation analysis are \texttt{LALInference} and
\texttt{Bilby}~\cite{1409.7215,1811.02042}, which offer multiple sampling
methods.  
  
%
%
Machine learning has featured prominently in many areas of \ac{GW} research
over the last few years. These techniques have shown to be particularly
promising in signal
detection~\cite{GEORGE201864,PhysRevLett.120.141103,GebKilParHarSch}, glitch
classification~\cite{0264-9381-34-6-064003}, earthquake
prediction~\cite{Coughlin_2017}, and to augment existing Bayesian sampling
methods~\cite{2012MNRAS.421..169G}. We also highlight recent developments
in \ac{GW} parameter estimation (independent to this work) where one- and
two-dimensional marginalised Bayesian posteriors are produced rapidly using
neural networks~\cite{2019arXiv190905966C}, and where normalised flows in
conjunction with \acp{CVAE} can reproduce Bayesian posteriors for a single
\ac{GW} detector case~\cite{PhysRevD.102.104057,Green_2021}. These methods, including the
one presented in this paper, are known as ``likelihood-free'' approaches in
which there is no requirement for explicit likelihood 
evaluation~\cite{Cranmer201912789}, only the need
to sample from the likelihood. Nor is it the case that pre-computed posterior
distributions are required in the training procedure.

%
%
Recently, a type of neural network known as \ac{CVAE} was shown to perform
exceptionally well when applied towards computational imaging
inference~\cite{1904.06264,NIPS2015_5775}, text to image
inference~\cite{1512.00570}, high-resolution synthetic image
generation~\cite{1612.00005} and the fitting of incomplete heterogeneous
data~\cite{1807.03653}. \acp{CVAE}, as part of the variational family of
inference techniques are ideally suited to the problem of function
approximation and have the potential to be significantly faster than existing
approaches. It is therefore this type of machine learning network that we
apply in the \ac{GW} case to accurately approximate the Bayesian posterior
$p(x|y)$, where $x$ represents the physical parameters that govern the
\ac{GW} signal, and are the quantities we are interested in inferring. The data
$y$ represents the noisy measurement containing the \ac{GW} signal and obtained
from a network of \ac{GW} detectors. 

%
%
The construction of a \ac{CVAE} begins with the definition of a quantity to be
minimised (referred to as a cost function). In our case we use the cross
entropy, defined as
\begin{align}\label{eq:cross_ent} 
H(p,r) &= -\int dx\, p(x|y) \log r_{\theta}(x|y) 
\end{align}
between the true posterior $p(x|y)$ and $r_{\theta}(x|y)$, the parametric
distribution that we will use neural networks to model and which we aim
to be equal to the true posterior. The parametric model is
constructed from a combination of 2 (encoder and decoder) neural networks
$r_{\theta_1}(z|y)$ and $r_{\theta_2}(x|y,z)$ where
\begin{align}\label{eq:latent_model}
r_{\theta}(x|y) = \int dz\,r_{\theta_1}(z|y)r_{\theta_2}(x|y,z).
\end{align}
In this case the $\theta$ subscripts represent sets of trainable neural network
parameters and the variable $z$ represents locations within a \emph{latent
space}. This latter object is typically a lower dimensional space within which
an encoder can represent the input data, and via marginalisation allows the
construction of a rich family of possible probability densities.

\begin{figure}
    \includegraphics[width=0.95\columnwidth]{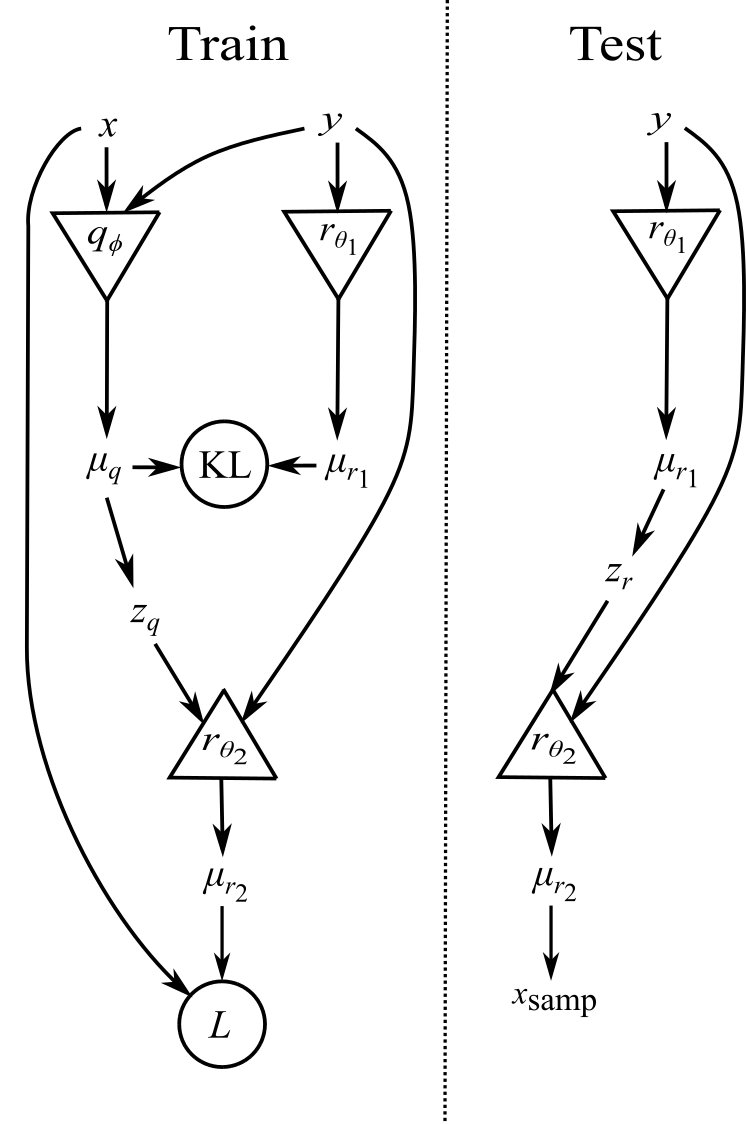}
    \caption{\label{fig:network_config} The configuration of the \ac{CVAE}
neural network. During training (left-hand side), a training set of noisy
\ac{GW} signals ($y$) and their corresponding true parameters ($x$) are given
as input to encoder network $q_{\phi}$, while only $y$ is given to
encoder network $r_{\theta_1}$. The \ac{KL}-divergence (Eq.~\ref{eq:kl})
is computed between the encoder output latent space representations
($\mu_q$ and $\mu_r$) forming one component of the total cost function. Samples
($z_q$) from the $q_{\phi}$ latent space representation are generated and
passed to the decoder network $r_{\theta_2}$ together with the original
input data $y$. The output of the decoder ($\mu_{r_2}$) describes a distribution in
the physical parameter space and the cost component $L$ is computed by
evaluating that distribution at the location of the original input $x$.
When performed in batches this scheme allows the computation of the total cost
function Eq.~\ref{eq:cost3}. After having trained the network and
therefore having minimised the cross-entropy $H$, we test (right-hand side)
using only the $r_{\theta_1}$ encoder and the $r_{\theta_2}$
decoder to produce samples ($x_{\text{samp}}$). These samples are drawn
from the distribution $r_{\theta}(x|y)$ (Eq.~\ref{eq:latent_model})
and accurately model the true posterior $p(x|y)$.}
\end{figure}

Starting from Eq.~\ref{eq:cross_ent} it is possible to derive a
computable bound for the cross-entropy that is reliant on the
$r_{\theta_1}$ and $r_{\theta_2}$ networks and a third ``recognition'' encoder
network $q_{\phi}(z|x,y)$ governed by the trainable parameter-set $\phi$. The
details of the derivation are described in the methods section and
in~\cite{1904.06264} but equate to an optimisation of the \ac{ELBO}. The
final form of the cross-entropy cost function is given by the bound

\begin{align}\label{eq:cost3} H \lesssim
\frac{1}{N}\sum_{n=1}^{N_{\text{b}}}&\Big[\overbrace{-\log
r_{\theta_{2}}(x_{n}|z_{n},y_{n})}^{L}\nonumber\\
&+\overbrace{\text{KL}\left[q_{\phi}(z|x_{n},y_{n})||r_{\theta_{1}}(z|y_{n})\right]}^{\text{KL}}\Big],
\end{align}
which is also represented graphically in Fig.~\ref{fig:network_config}.
The cost function is composed of 2 terms, the ``reconstruction'' cost $L$ which
is a measure of how well the decoder network $r_{\theta_2}$ predicts the true
signal parameters $x$, and the \ac{KL}-divergence cost that measures the
similarity between the distributions modelled by the $r_{\theta_1}$ and
$q_{\phi}$ encoder networks. In practice, for each iteration of the training
procedure, the integrations over $x,y$ and $z$ are approximated by a sum over a
batch of $N_{\text{b}}$ draws from the user defined prior $p(x)$, the known
likelihood $p(y|x)$, and the recognition function $q_{\phi}(z|,x,y)$. Details
of the training procedure are given in the methods section.  

%
%
The implementation of the \ac{CVAE} that we employ in this letter has a
number of specific features that were included in order to tailor the analysis to
\ac{GW} signals. The details of these enhancements are described in the Methods
section but in summary, the primary modifications are as follows, 1) Physically 
appropriate output decoder distributions are used for each output parameter: 
von Mises-Fisher distribution on the sky location parameters, von Mises 
distributions on all parameters with cyclic prior bounds,  and truncated Gaussians for
parameters with defined prior bounds. 2) Each of the
functions $r_{\theta_1},r_{\theta_2}$, and $q_{\phi}$ are modelled using deep
convolutional neural networks with multi-detector time-series represented as
independent input channels. 3) The $r_{\theta_1}$ encoder models an $M=32$ component
Gaussian mixture model within the $n_{z}=15$ dimensional latent space in order
to capture the corresponding typical multi-modal nature of \ac{GW} posterior
distributions. 4.) All cyclic parameters are represented as points in an abstract 
2D plane.

%
%

%
%
%
%
We present results on $250$ multi-detector \ac{GW} test \ac{BBH}
waveforms in simulated advanced detector noise~\cite{aligo_noisecurves}
from the LIGO Hanford, Livingston and Virgo detectors. We compare between
variants of the existing Bayesian approaches and our \ac{CVAE} implementation
which we call \texttt{VItamin}~\footnote{\url{https://github.com/hagabbar/VItamin.git}}.
 Posteriors produced by the \texttt{Bilby}
inference library~\cite{1811.02042} are used as a benchmark in order to assess
the efficiency and quality of our machine learning approach with the existing
methods for posterior sampling.

%
%
For the benchmark analysis we assume that 14 parameters are
unknown: the component masses
$m_1,m_2$, the luminosity distance $d_{\text{L}}$, the sky position
$\alpha,\delta$, the binary inclination $\Theta_{jn}$, the \ac{GW} polarisation
angle ${\psi}$, the time of coalescence $t_{0}$, and the spin parameters $a_1,a_2,
\Theta_1,\Theta_2,\phi_{12},\phi_{jl}$.  We do not include phase $\phi_0$ in our 
results 
because we apply phase marginalisation to all Bayesian samplers since 
this 
improves overall stability and runtime~\cite{1811.02042}.For each parameter 
we use a uniform prior with the exception of 
the declination, inclination, and tilt angle parameters for 
which we use priors uniform in $\cos\delta$, $\sin\Theta_{jn}$, $\sin\Theta_1$, 
and $\sin\Theta_2$ respectively.  The prior on the component masses are conditional, 
such that $m_1 > m_2$.
The corresponding prior ranges
are defined in Table~\ref{tab:prior_ranges} and result in a  \ac{SNR}
distribution that has a median value of $\text{SNR}\approx 9$ and ranging between 0 and 75.
We use a sampling frequency of $1024$~Hz, a time-series duration of 1 second, and
the waveform model used is \texttt{IMRPhenomPv2}~\cite{1809.10113} with a
minimum cutoff frequency of $20$Hz. For each input test waveform we run the
benchmark analysis using multiple sampling algorithms available within
\texttt{Bilby}. For each run and sampler we extract on the order of 8000
samples from the posterior on the 14 physical parameters.  

%
%
The \texttt{VItamin} training process uses as input $10^{7}$ whitened
waveforms corresponding to parameters drawn from the same priors as assumed for
the benchmark analysis. The waveforms are also of identical duration, sampling
frequency, and use the same waveform model as in the benchmark analysis.
The signals are whitened\footnote{The whitening is used primarily to
scale the input to a magnitude range more suitable to neural networks. The
\emph{true} \ac{PSD} does not have to be used for whitening, but training data
and test data must be contain signals that share the same \ac{PSD}.}
using the same advanced detector \acp{PSD}~\cite{aligo_noisecurves} as
assumed in the benchmark analysis. When each whitened waveform is placed
within a training batch it is given a unique detector Gaussian noise
realisation (after signal whitening this is simply zero mean, unit
variance Gaussian noise). The \texttt{VItamin} posterior results are produced by
passing each of our $250$ whitened noisy testing set of \ac{GW} waveforms
as input into the testing path of the pre-trained
\ac{CVAE} (Fig.~\ref{fig:network_config}). For each input waveform we sample until we
have generated $8000$ posterior samples on 15 physical parameters, collectively denoted 
here as $x$.  Comparison results between \texttt{VItamin} and Bayesian samplers 
do not use $\phi_0$ since it is marginalised out in the Bayesian sampler inference 
process.

%
%
\begin{figure*}
    \includegraphics[width=\textwidth]{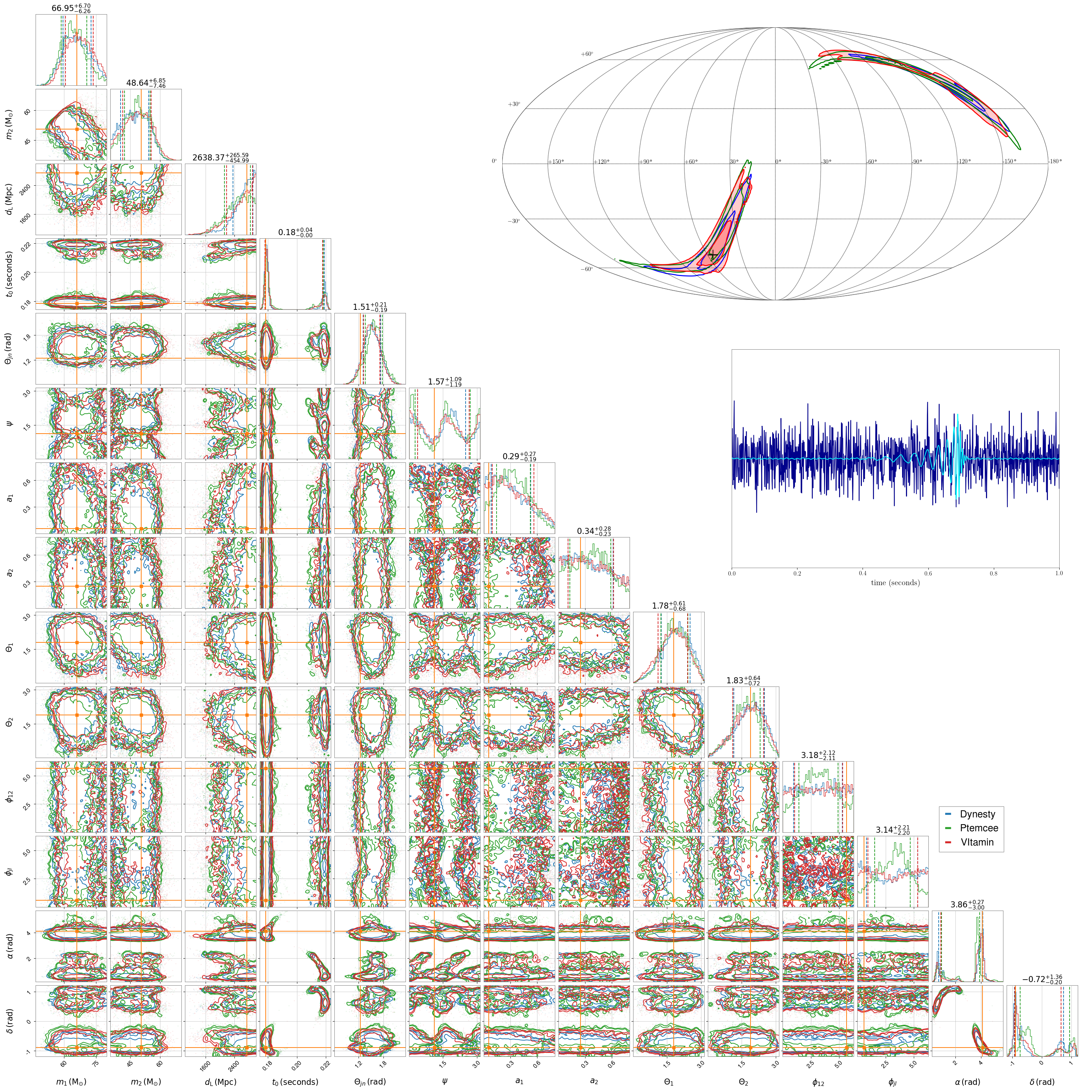}
    \caption{\label{fig:corner_plot} Corner plot showing one and two-dimensional
marginalised posterior distributions on the \ac{GW} parameters for one
example test dataset. Red contours represent the
two-dimensional joint posteriors obtained from \texttt{VItamin} and
blue and green contours are the corresponding posteriors
output from our benchmark analyses (using the \texttt{Dynesty} and \texttt{ptemcee}
samplers within \texttt{Bilby}). In each case, the contour boundaries enclose
$68,90$ and $95\%$ probability. One dimensional histograms of the posterior
distribution for each parameter from both methods are plotted along the
diagonal. Vertical dashed lines in the one dimensional plots are representative of the
 $5\%$ --- $95\%$ symmetric confidence bounds of the 3 sampler 1 dimensional posteriors. Orange vertical and horizontal lines denote the true parameter
values of the simulated signal. At the top of the figure we include a Mollweide
projection of the sky location posteriors from all three analyses. All results
presented in this letter correspond to a three-detector configuration but for
clarity we only plot the H1 whitened noisy time-series $y$ and the noise-free
whitened signal (in blue and cyan respectively) to the right of the figure. The
test signal was simulated with an optimal multi-detector signal-to-noise ratio of
14.3.} 
\end{figure*}

%
%
We can immediately illustrate the accuracy of our machine learning predictions
by directly plotting 2 and one-dimensional marginalised posteriors generated
using the output samples from our \texttt{VItamin} and \texttt{Bilby}
approaches superimposed on each other. We show this for one example test
dataset in Fig.~\ref{fig:corner_plot} where strong agreement between the
\texttt{Bilby} sampler \texttt{Dynesty} in blue,  and the
\ac{CVAE} (red) is clear. It is also evident that whilst we refer to the
\texttt{Bilby} sampler results as benchmark cases, different existing samplers
do not perfectly agree with each other (i.e. \texttt{ptemcee} in green) despite using 
expert recommended sampler settings shown in Tab.~\ref{Tab:sampler_params}.  
For each of our 250 test cases we see
reasonable levels of agreement between pairs of benchmark samplers \emph{and}
between any benchmark sampler and our \ac{CVAE} results. 

%
%
Figures~\ref{fig:pp_plot} and \ref{fig:kl_results} (see the Methods section)
show the results of 2 statistical tests (the \ac{PP} plot test and
\ac{JS}-divergence tests) performed on the entire test dataset and between all
samplers (\texttt{Dynesty}, \texttt{ptemcee}, \texttt{CPNest}, \texttt{emcee}, and \texttt{VItamin}). In
both tests the quality of the \texttt{VItamin} results are
 reasonably consistent with the
benchmark samplers. The \ac{PP} plot results specifically indicate that the
Bayesian one-dimensional marginalised posteriors from each approach are
self-consistent from a frequentist perspective (e.g., the true values lie
within the $X\%$ confidence interval for $X\%$ of the test cases). The second
test computes the distribution of \ac{JS}-divergences between posteriors
conditioned on the same test data $y$ from the ``gold standard'' benchmark 
sampler \texttt{Dynesty} and every other benchmark sampler (including 
\texttt{VItamin}).  This measure of ``
distribution similarity'' shows that the results of \texttt{VItamin} vs. 
\texttt{Dynesty} generally lie between those of  \texttt{CPNest} vs. 
\texttt{Dynesty} and \texttt{ptemcee} vs. \texttt{Dynesty}.

%
%
The dominating computational cost of running \texttt{VItamin} lies in the
training time, which takes on the order of 7 days to complete. We
stress that once trained, there is no need to retrain the network unless the
user wishes to use different priors $p(x)$ or assume different noise
characteristics. The speed at which posterior samples are generated for all
samplers used, including \texttt{VItamin}, is shown in Table~\ref{Tab:speed}.
Run-time for the benchmark samplers is defined as the time to complete their
analyses when configured using the parameter choices defined in
Table~\ref{Tab:sampler_params}. For \texttt{VItamin}, this time is defined as
the total time to produce $8000$ samples. For our test case of \ac{BBH} signals
\texttt{VItamin} produces samples from the posterior at a rate which is $\sim
6$ orders of magnitude faster than our benchmark analyses using current
inference techniques. 


%
%
\begin{table}
\centering
\caption{Durations required to produce samples from each of
the different posterior sampling approaches.}
\begin{tabular}[t]{lcccc} 
\toprule
\multirow{2}{*}{sampler} & \multicolumn{3}{c}{run time (seconds)} & \multirow{2}{*}{ratio
$\displaystyle\frac{\tau_{\text{VItamin}}}{\tau_{X}}$} \\
& min & max & median & \\
\hline
\texttt{Dynesty}\footnote{The benchmark samplers all produced
on the order of 8000 samples dependent on the default sampling parameters
used.}~\cite{dynesty} & 21564 & 261268 & 45607
\footnote{We note that there are a growing number of specialised
techniques~\cite{2016PhRvD..94d4031S,2019PhRvD..99h4026W,2019PhRvD.100d3030T,PhysRevD.92.023002}
designed to speed up traditional sampling algorithms that could be used to
reduce the runtimes quoted here by approximately 1-2 orders of magnitude.}
& $2.2\times 10^{-6}$ \\
\texttt{emcee}~\cite{emcee} & 16712 & 39930 & 19821 & $5.1\times 10^{-6}$ \\
\texttt{ptemcee}~\cite{ptemcee} & 2392 & 501632 & 41151.0 & $2.4\times 10^{-6}$ \\
\texttt{CPNest}~\cite{cpnest} & 10309 & 437008 & 83807 & $1.2\times 10^{-6}$ \\
\texttt{VItamin}\footnote{For the \texttt{VItamin} sampler $8000$ samples are
produced as representative of a typical posterior. The run time is independent
of the signal content in the data and is therefore constant for all test cases.} & \multicolumn{3}{c}{\bm{$1\times10^{-1}$}} & 1 \\
\botrule
\end{tabular}
\label{Tab:speed}
\end{table}

%
%

%
%
%
In this letter we have demonstrated that we are able to reproduce, to a high
degree of accuracy, Bayesian posterior probability distributions generated
through machine learning. This is accomplished using a \ac{CVAE} trained on
simulated \ac{GW} signals and does not require the input of precomputed
posterior estimates. We have demonstrated that our neural network model, which
when trained, can produce complete and accurate posterior estimates in
a fraction of a second, achieves the same quality of results as the
trusted benchmark analyses used within the \ac{LVC}.

%
%
The significance of our results is most evident in the orders of magnitude
increase in speed over existing algorithms. We have demonstrated the approach
using \ac{BBH} signals but with additional work to increase sample rate
and signal duration, the method can also be extended for application to
signals from \ac{BNS} mergers (e.g., GW170817~\cite{PhysRevLett.119.161101},
and GW190425~\cite{2020ApJ...892L...3A}) and \ac{NSBH}~\cite{Abbott_2021} systems where
improved low-latency alerts will be especially pertinent. By using our
approach, parameter estimation speed will no longer be limiting
factor\footnote{A complete low-latency pipeline includes a number of
steps. The process of \ac{GW} data acquisition is followed by the transfer of
data. There is then the corresponding candidate event identification,
parameter estimation analysis, and the subsequent communication of results to
the \ac{EM} astronomy community after which there are physical aspects such as
slewing observing instruments to the correct pointing.} in observing the prompt
\ac{EM} emission expected on shorter time scales than is achievable with
existing \ac{LVC} analysis tools such as Bayestar~\cite{2016PhRvD..93b4013S}.

%
%
The predicted number of future detections of \ac{BNS} mergers ($\sim
180$~\cite{2018LRR....21....3A}) will severely strain the \ac{GW} community's
current computational resources using existing Bayesian methods. We anticipate
that future iterations of our approach will provide full-parameter estimation
on all classes of \ac{CBC} signals in approximately 1 second on single
\acp{GPU}. Our trained network is also modular, and can be shared and used
easily by any user to produce results. The specific analysis described in this
letter assumes a uniform prior on the signal parameters. However, this is a
choice and the network can be trained with any prior the user demands, or users
can cheaply resample accordingly from the output of the network trained on the
uniform prior. We also note that our method will be invaluable for population
studies since populations may now be generated and analysed in a fully-Bayesian
manner on a vastly reduced time scale. 

%
%
For \ac{BBH} signals, \ac{GW} data is usually sampled at $1$---$4$ kHz
dependent upon the mass of the binary. We have chosen to use the noticeably low
sampling rate of 1024Hz in order to decrease the computational time
required to develop our approach and the computational burden of computing our
250 benchmark analyses for each of 4 benchmark samplers.  We have found that 
increasing the sampling frequency of our input comes at the cost of an increase 
in training time and a similar increase on the \ac{GPU} memory
requirement. We note that with the exception of requiring one-dimensional
convolutional layers and an increase in the amount of training data to
efficiently deal with a multi-detector analysis, the network complexity has not
increased 
with the dimensionality of the physical parameter space nor with the
sampling rate of the input data. We therefore do not anticipate that extending
the parameter space to lower masses will be problematic. 

%
%
In reality, \ac{GW} detectors are affected by non-Gaussian noise artefacts and
time-dependent variation in the detector noise \ac{PSD}. Existing methods
incorporate a parameterised \ac{PSD} estimation into their
inference~\cite{2015PhRvD..91h4034L}. To account for these and to exploit the
``likelihood-free'' nature of the \ac{CVAE} approach, we could re-train our
network at regular intervals using samples of real detector noise (preferably
recent examples to best reflect the state of the detectors). In this case
we could also apply transfer learning to speed up each training instance based
on the previously trained network state.  Alternatively, since the \ac{PSD} is
an estimated quantity, we could marginalise over its uncertainty by providing
training data whitened by samples drawn from a distribution of possible
\acp{PSD}. Our work can naturally be extended to include the full range of
\ac{CBC} signal types but also to any and all other parameterised \ac{GW}
signals and to analyses of \ac{GW} data beyond that of ground based
experiments. Given the abundant benefits of this method, we hope that a variant
of this of approach will form the basis for future \ac{GW} parameter
estimation.

%
%

%
%
\section{Acknowledgements.}
We would like to acknowledge valuable input from the LIGO-Virgo Collaboration,
specifically from Will Farr, Tom Dent, Jonah Kanner, Alex Nitz, Colin Capano 
and the parameter estimation 
and machine-learning working groups. We would additionally like to thank Szabi 
Marka for posing this challenge to us and the journal referees for their 
helpful and constructive comments. We thank Nvidia for the generous donation 
of a Tesla V-100 GPU used in addition to \ac{LVC} computational resources. 
The authors also gratefully acknowledge the Science and Technology Facilities 
Council of the United Kingdom. CM and SH are supported by the Science and 
Technology Research Council (grant No.~ST/~L000946/1) and the European 
Cooperation in Science and Technology (COST) action CA17137. FT acknowledges 
support from Amazon Research and  EPSRC grant EP/M01326X/1, and RM-S EPSRC 
grants EP/M01326X/1, EP/T00097X/1 and EP/R018634/1. 

%
%


\section{addendum}
 \subsection{Competing Interests} 
    The authors declare that they have no competing financial interests.
 \subsection{Correspondence} Correspondence and requests for materials should be addressed to Hunter Gabbard~(email: hunter.gabbard@gmail.com).

%
%
\section{Methods}\label{sec:methods}
%

%
%
A \ac{CVAE} is a form of variational autoencoder that is
conditioned on an observation, where in our case the observation is a
one-dimensional \ac{GW} time-series signal $y$. The autoencoders from which
variational autoencoders are derived are typically used for problems involving
image reconstruction and/or dimensionality reduction. They perform a regression
task whereby the autoencoder attempts to predict its own given input (model the
identity function) through a ``bottleneck layer'' --- a limited  and therefore
distilled representation of the input parameter space. An autoencoder is
composed of two neural networks, an encoder and a
decoder~\cite{gallinari1987memoires}.  The encoder network takes as input a
vector, where the number of dimensions is a fixed number predefined by the
user. The encoder converts the input vector into a (typically) lower
dimensional space, referred to as the {\it{latent space}}. A representation of
the data in the latent space is passed to the decoder network which generates a
reconstruction of the original input data to the encoder network. Through
training, the two sub-networks learn how to efficiently represent a dataset
within a lower dimensional latent space which will take on the most important
properties of the input training data. In this way, the data can be compressed
with little loss of fidelity. Additionally, the decoder simultaneously learns
to decode the latent space representation and reconstruct that data back to its
original form (the input data).

%
%
The primary difference between a variational autoencoder~\cite{1812.04405} and
an autoencoder concerns the method by which locations within the latent space
are produced. In our variant of the variational autoencoder, the output of the
encoder is interpreted as a set of parameters governing statistical
distributions. In proceeding to the decoder network, samples from the latent
space ($z$) are randomly drawn from these distributions and fed into the
decoder, therefore adding an element of variation into the process. A
particular input can then have a range of possible outputs. Any trainable
network architectures can be used in both the decoder and the encoder networks
and within \texttt{VItamin} we use deep convolutional neural networks in both
cases.

\subsection{Cost function derivation}
%
%
We will now derive the cost function and the corresponding network structure
and we begin with the statement defining the aim of the analysis. We wish to
obtain a function that reproduces the posterior distribution (the probability
of our physical parameters $x$ given some measured data $y$). The cross-entropy
between 2 distributions is defined in Eq.~\ref{eq:cross_ent} where we have made
the distributions explicitly conditional on $y$ (our measurement). In this case
$p(x|y)$ is the target distribution (the true posterior) and $r_{\theta}(x|y)$
is the parametric distribution that we will use neural networks to construct.
The variable $\theta$ represents the trainable neural network parameters. 

%
%
The cross-entropy is minimised when $p(x|y)=r_{\theta}(x|y)$ and so by
minimising
\begin{align}\label{eq:cost1}
H &= -\text{E}_{p(y)}\left[\int dx\,p(x|y) \log r_{\theta}(x|y)\right],
\end{align}
where $\text{E}_{p(y)}[\cdot]$ indicates the expectation value over the
distribution of measurements $y$, we therefore make the parametric distribution
as similar as possible to the target for all possible measurements $y$.

%
%
Converting the expectation value into an integral over $y$ weighted by $p(y)$
and applying Bayes' theorem we obtain
\begin{align}\label{eq:cost1}
H &= -\int dx\,p(x)\int dy\,p(y|x)\log r_{\theta}(x|y)
\end{align}
where $p(x)$ is the prior distribution on the physical parameters $x$, and
$p(y|x)$ is the likelihood of $x$ (the probability of measuring the data $y$
given the parameters $x$). 

%
%
The \ac{CVAE} network outlined in Fig.~\ref{fig:network_config} makes use of a
conditional latent variable model and our parametric model is constructed from
the product of 2 separate distributions marginalised over the latent space as
defined in Eq.~\ref{eq:latent_model}. We have used $\theta_{1}$ and
$\theta_{2}$ to indicate that the 2 separate networks modelling these
distributions will be trained on these parameter sets respectively. The
encoder $r_{\theta_1}(z|y)$ takes as input the data $y$ and outputs parameters
that describe a probability distribution within the latent space. The decoder
$r_{\theta_2}(x|z,y)$ takes as input a single location $z$ within the latent
space together with the data $y$ and outputs sets of parameters describing a
probability distribution in the physical parameter space. 

%
%
One could be forgiven for thinking that by setting up networks that simply aim
to minimise $H$ over the $\theta_{1}$ and $\theta_{2}$ would be enough to solve
this problem. However, as shown in~\cite{NIPS2015_5775} this is an intractable
problem and a network cannot be trained directly to do this. Instead we
introduce a recognition function $q_{\phi}(z|x,y)$, modelled by an
additional neural network and governed by the trainable network parameters
$\phi$, that will be used to derive an \ac{ELBO}.

%
%
Let us first define the \ac{KL}-divergence between the recognition function and
the distribution $r_{\theta}(z|x,y)$ as 
\begin{align}\label{eq:kl}
\text{KL}&\left[q_{\phi}(z|x,y)||r_{\theta}(z|x,y)\right] = \\
&\int dz\,q_{\phi}(z|x,y)
\log\left(\frac{q_{\phi}(z|x,y)}{r_{\theta}(z|x,y)}\right),\nonumber
\end{align}
from which it can be shown that
\begin{align}\label{eq:elbo1}
\log r_{\theta}(x|y) &= \text{ELBO} + \text{KL}\left[q_{\phi}(z|x,y)||r_{\theta}(z|x,y)\right],
\end{align}
where the \ac{ELBO} is given by
\begin{align}\label{eq:elbo2}
\text{ELBO} &= \int dz\,
q_{\phi}(z|x,y)\log\left(\frac{r_{\theta_{2}}(x|y,z)r_{\theta_{1}}(z|y)}{q_{\phi}(z|x,y)}\right).
\end{align}
It is so-named since the \ac{KL}-divergence has a minimum of zero and cannot
be negative. Therefore, if we were to find a $q_{\phi}(z|x,y)$ function (optimised on
$\phi$) that minimised the \ac{KL}-divergence defined in Eq.~\ref{eq:kl} then we can state that
\begin{align}
\log r_{\theta}(x|y) &\geq \text{ELBO}.
\end{align}
After some further manipulation of Eq.~\ref{eq:elbo2} we find that
\begin{align}\label{eq:logr}
\log r_{\theta}(x|y) \geq  &\text{E}_{q_{\phi}(z|x,y)}\left[\log
r_{\theta_{2}}(x|z,y)\right] \nonumber\\
&-\text{KL}\left[q_{\phi}(z|x,y)||r_{\theta_{1}}(z|y)\right].
\end{align}
We can now substitute this inequality into our cost function as defined by
Eq.~\ref{eq:cost1} to obtain
\begin{align}\label{eq:cost2}
H \leq  -\int dx\, p(x)&\int dy \,p(y|x)
\Big[\text{E}_{q_{\phi}(z|x,y)}\left[\log r_{\theta_{2}}(x|z,y)\right]
\nonumber\\
&-\text{KL}\left[q_{\phi}(z|x,y)||r_{\theta_{1}}(z|y)\right]\Big],  
\end{align}
which can in practice be approximated as a stochastic integral over draws of
$x$ from the prior, $y$ from the likelihood function $p(y|x)$, and from the
recognition function, giving us Eq.~\ref{eq:cost3}, the actual function
evaluated within the training procedure. In standard sampling algorithms it is
required that the likelihood is calculated explicitly during the exploration of
the parameter space and hence an analytic noise and signal model must be
assumed. For a \ac{CVAE} implementation we are required only to sample from
the likelihood distribution, i.e., generate simulated noisy measurements given a
set of signal parameters. This gives us the option of avoiding the assumption
of detector noise Gaussianity in the future by training the \ac{CVAE} using ``real''
non-Gaussian detector noise.

%
%
\begin{figure}
    \includegraphics[width=\columnwidth]{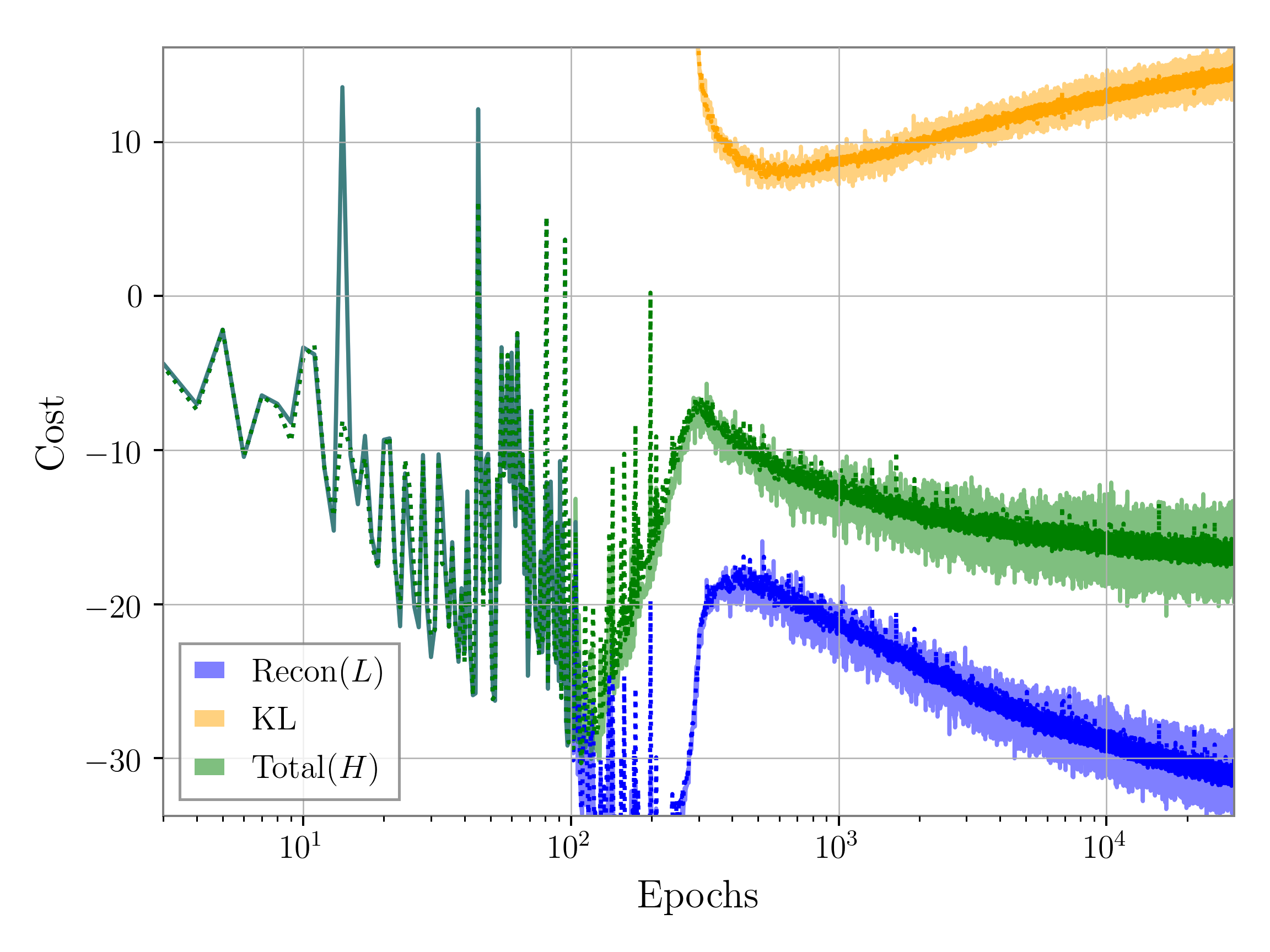}
\caption{\label{fig:loss_log} The cost as a function of training epoch.
We show the total cost function (green) together with its component parts:
the \ac{KL}-divergence component (orange) and the reconstruction component
(blue) which are simply summed to obtain the total. The dark curves correspond
to the cost computed on each batch of training data and the lighter curves
represent the cost when computed on independent validation data. The close
agreement between training and validation cost values indicates that the
network is not overfitting to the training data. The change in behavior of the
cost between $10^2$ and $3\times10^2$ epochs is a consequence of gradually
introducing the \ac{KL} cost term contribution via an annealing process.} 
\end{figure}

%
%
\begin{table}
\centering
\caption{The uniform prior boundaries and fixed parameter values used on the \ac{BBH} signal parameters for the benchmark
and the \ac{CVAE} analyses.}
\begin{tabular}[t]{lcccc}
\toprule
Parameter name & symbol & min & max & units \\
\hline
mass 1 & $m_1$ & 35 & 80 & solar masses \\
mass 2 & $m_2$\footnote{Additionally $m_2$ is constrained such that
$m_{2}<m_{1}$.} & 35 & 80 & solar masses \\
luminosity distance & $d_{\text{L}}$ & 1 & 3 & Gpc \\
time of coalescence & $t_{0}$ & 0.65 & 0.85 & seconds \\
phase at coalescence & $\phi_{0}$ & 0 & $2\pi$ & radians \\
right ascension & $\alpha$ & 0 & $2\pi$ & radians \\
declination & $\delta$ & $-\pi/2$ & $\pi/2$ & radians \\
inclination & $\Theta_{jn}$ & 0 & $\pi$ & radians \\
polarisation & $\psi$ & 0 & $\pi$ & radians \\
spin magnitude 1 & $a_1$ & 0 & 0.8 & - \\
spin magnitude 2 & $a_2$ & 0 & 0.8 & - \\
tilt angle 1 & $\Theta_1$ & 0 & $\pi$ & radians \\
tilt angle 2 & $\Theta_2$ & 0 & $\pi$ & radians \\
azimuthal angle & $\phi_{12}$ & 0 & $2\pi$ & radians \\
azimuthal position & $\phi_{jl}$ & 0 & $2\pi$ & radians \\
\hline
epoch & \multicolumn{3}{c}{1126259642} & GPS time \\
detector network & \multicolumn{3}{c}{LIGO H1,L1, \& Virgo V1} & - \\
\botrule
\end{tabular}
\label{tab:prior_ranges}
\end{table}

\subsection{Network design}\label{sec:network_design}
%
%

%
%
The \ac{CVAE} network outlined in Fig.~\ref{fig:network_config} is constructed from
the 3 separate neural networks modelling the encoder and decoder distributions
$r_{\theta_1}$ and $r_{\theta_2}$ as well as the recognition function
$q_{\phi}$. Each of these components is a deep convolutional network
consisting of a series of one-dimensional convolutional layers followed by a series of
fully-connected layers, where convolutional layers are shared between all 3 networks. The 
details of each network structure are given in
Table~\ref{Tab:network_design} where we indicate the activations used and additional
dropout and batch-normalisation layers.

%
%
\begin{table*}
\centering
\caption{The \texttt{VItamin} network hyper-parameters. Dashed lines ``---'' indicate that 
convolutional layers are shared between all 3 networks. }
\begin{tabular}[t]{l|ccc}
\toprule
\backslashbox{Layer}{Network} & $r_{\theta_1}(z|y)$ & $r_{\theta_2}(x|y,z)$ & $q_{\phi}(z|x,y)$ \\
\hline
\multirow{2}{*}{Input $y$} & \multirow{2}{*}{[1024,3]\footnote{The shape of the
data [one-dimensional dataset length, No. channels].}} &
\multirow{2}{*}{[1024,3]} & \multirow{2}{*}{[1024,3]} \\
& & & \\
\hline
\multirow{2}{*}{Layer 1} & conv(64,3,96)\footnote{one-dimensional
convolutional filter with arguments (filter size, No. channels, No. filters).} & --- & --- \\
& L2Reg(0.001)\footnote{L2 regularization funciton applied to the kernel weights 
matrix.} & --- & --- \\
& act\footnote{The activation function used.}=LeakyReLU & --- & --- \\
\hline
\multirow{3}{*}{Layer 2} & conv(32,96,96) & --- &
--- \\
& stride(4)\footnote{Striding layer with arguments (stride
length).} & --- & --- \\
& L2Reg(0.001) & --- & --- \\
& act=LeakyReLU & --- & --- \\
\hline
\multirow{2}{*}{Layer 3} & conv(32,96,96) & --- &
--- \\
& L2Reg(0.001) & --- & --- \\
& act=LeakyReLU & --- & --- \\
\hline
\multirow{2}{*}{Layer 4} & conv(16,96,96) & --- &
--- \\
& stride(2) & --- & --- \\
& L2Reg(0.001) & --- & --- \\
& act=LeakyReLU & --- & --- \\
\hline
\multirow{2}{*}{Layer 5} & conv(16,96,96) & --- &
--- \\
& L2Reg(0.001) & --- & --- \\
& act=LeakyReLU & --- & --- \\
\hline
\multirow{2}{*}{Layer 6} & conv(16,96,96) & --- &
--- \\
& stride(2) & --- & --- \\
& L2Reg(0.001) & --- & --- \\
& act=LeakyReLU & --- & --- \\
\hline
\multirow{2}{*}{Input $z,x$} & \multirow{2}{*}{flatten\footnote{Take the multi-channel output of the previous layer and
reshape it into a one-dimensional vector.}$\rightarrow$[6144]} &
flatten$\rightarrow$[6144] & flatten$\rightarrow$[6144] \\
& & append\footnote{Append the argument to the current dataset.}($z$)$\rightarrow$[6159] & append($x$)$\rightarrow$[6159] \\
\hline
\multirow{3}{*}{Layer 7} & 
\multirow{3}{*}{
\begin{tabular}[t]{c}
FC(6159,4096)\footnote{Fully
connected layer with arguments (input size, output size).}\\
act=LeakyReLU \\
\end{tabular}
} & 
\multirow{3}{*}{
\begin{tabular}[t]{c}
FC(6159,4096) \\
act=LeakyReLU \\
\end{tabular}
} &
\multirow{3}{*}{
\begin{tabular}[t]{c}
FC(6159,4096) \\
act=LeakyReLU \\
\end{tabular}
} \\
& & & \\
& & & \\
\hline
\multirow{3}{*}{Layer 8} & 
\multirow{3}{*}{
\begin{tabular}[t]{c}
FC(4096,2048)\\
act=LeakyReLU \\
\end{tabular}
} & 
\multirow{3}{*}{
\begin{tabular}[t]{c}
FC(4096,2048) \\
act=LeakyReLU \\
\end{tabular}
} &
\multirow{3}{*}{
\begin{tabular}[t]{c}
FC(4096,2048) \\
act=LeakyReLU \\
\end{tabular}
} \\
& & & \\
& & & \\
\hline
\multirow{3}{*}{Layer 9} & 
\multirow{3}{*}{
\begin{tabular}[t]{c}
FC(2048,1024)\\
act=LeakyReLU \\
\end{tabular}
} & 
\multirow{3}{*}{
\begin{tabular}[t]{c}
FC(2048,1024)\\
act=LeakyReLU \\
\end{tabular}
} &
\multirow{3}{*}{
\begin{tabular}[t]{c}
FC(2048,1024) \\
act=LeakyReLU \\
\end{tabular}
} \\
& & & \\
& & & \\
\hline
\multirow{4}{*}{Layer 10} & 
\multirow{4}{*}{
\begin{tabular}[t]{c}
FC(1024,960) \\
act=None \\
output=$\mu_{r_1}$ \\
$\rightarrow$[15,32,2]\footnote{The $r_{\theta_1}$ output has size
[latent space dimension, No. modes, No. parameters defining each
component per dimension].} \\
\end{tabular}
} & \multirow{4}{*}{
\begin{tabular}[t]{c}
FC(1024,30) \\
act=(Sigmoid,-ReLU)\footnote{Different activations are used for different
parameters. For the scaled parameter means we use
sigmoids and for log-variances we use negative ReLU functions.} \\
output=$\mu_{r_2}$ \\
$\rightarrow$[19,2]\footnote{The $r_{\theta_2}$ output has size [physical space dimension+additional cyclic dimensions, No. parameters defining
the distribution per dimension].  The addtional cyclic dimensions account for the 2 parameters 
each cyclic parameter is represented by in the abstract 2D plane.} \\
\end{tabular}
} &
\multirow{4}{*}{
\begin{tabular}[t]{c}
FC(1024,30) \\
act=None \\
output=$\mu_{q}$ \\
$\rightarrow$[15,2]\footnote{The $q_{\phi}$ output has size [latent space
dimension, No. parameters defining the distribution per dimension].} \\
\end{tabular}
} \\
& & & \\
& & & \\
& & & \\
\botrule
\end{tabular}
\label{Tab:network_design}
\end{table*}

%
%
The $r_{\theta_1}$ network takes the input time-series data $y$ in the form of
multi-channel 1-dimensional vectors where channels represent different \ac{GW}
detectors. After passing through a series of one-dimensional convolutional and fully connected
layers, the output then defines the parameters of a $n_z$-dimensional
(diagonal) Gaussian mixture model in the latent space. We label these 
parameters as $\mu_{r_1}$ containing $n_z\times M$ means 
and log-covariances, where $M=32$ mixture component weights 
and $n_z = 15$. The motivation for using this
mixture model representation comes from the multi-modal nature of \ac{GW}
posterior distributions. The encoder network can use this flexibility to
represent the $y$ time-series data as belonging to multiple possible latent
space regions. 

%
%
The recognition function network $q_{\phi}$ is very similar to the
$r_{\theta_1}$ network with only 2 differences. The network takes as input the
$y$ time-series and the true signal parameters $x$, however, only the $y$ data
is passed through the one-dimensional convolutional layers. Only after the
final convolutional layer where the output is flattened is the $x$ data
appended. It is then this compound time-series data ``feature-space'' and true
signal parameters that are processed using the remaining fully-connected
layers. The second difference is that the output of the network defines a
\emph{single-modal} (diagonal) $n_z$-dimensional Gaussian. We label these
parameters as $\mu_{q}$ containing $n_z=15$ means and log-covariances. The
rationale behind this choice is that since the $q_{\phi}$ distribution is
conditional on the true signal parameters, there should be no ambiguity as to
which mode in the latent space that a particular time-series belongs to.      

%
%
The $r_{\theta_2}$ output represents the parameters ($\mu_{r_2}$) that govern an
$n_x$-dimensional distribution in the physical parameter space and we have
carefully chosen appropriate distributions for each of the physical parameters.
For the two component masses, luminosity distance, the binary inclination, the time of
coalescence, and spin parameters $a_1,a_2,\Theta_1,\Theta_2$ we have 
adopted truncated Gaussian distributions where the
truncation occurs at the predefined prior boundaries of the respective
parameter space dimensions. 
Independent von Mises distributions are applied for the polarization angle,  phase, 
and spin parameters $\phi_{12},\phi_{jl}$ in order to capture the periodic 
nature of these parameters. Finally, we use the von Mises-Fisher distribution
to model the right ascension and declination (sky) parameters.  Each cyclic parameter 
is represented as two predicted numbers in an abstract 2D plane, whereby the angle 
between the two numbers is representative of each cyclic parameter value.  This 2D 
representation is beneficial because there are no boundaries in this space and thus 
it is easier for the neural network to produce predictions which lie on the wrapped edges 
of the periodic bounds of each cyclic parameter.

%
%
We additionally reparameterise phase and the polarisation angle.   
This is done in order to simplify the search space for 
the neural network and was partly influenced by the work of~\cite{10.1093/mnras/stv1584}. 
This is accomplished by representing $\psi$ and $\phi_ 0$ as two new parameters 
$\psi^{'}$ and $X$. $X$ is given as the modulus of $(\psi+\phi_0)/(\pi)$ and 
$\psi^{'}$ is given as the modulus of $(\psi)/(\pi/2)$. The 
parameterisation given above acts to effectively reduce the number 
of modes seen by the network, thus making the search space simpler. After training, the network 
will output posterior samples in the $X,\psi^{'}$ space  and we 
must then convert back to the original $\psi$ and $\phi_ 0$ space to produce our final 
set of posterior samples.  Whilst this parameterisation is completely acceptable for 
spinning and precessing waveforms, it is not appropriate when considering higher order 
modes since the $\phi_0,\psi$ degeneracy is broken.

%
%


\subsection{Training procedure}\label{app:training_procedure}
%
%
Our cost function is composed of 3 probability distributions modelled by
neural networks with well defined inputs and outputs where the mapping of
those inputs to outputs is governed by the parameter sets
$\theta_{1},\theta_{2}$ and $\phi$. These parameters are the weights and biases
of 3 neural networks acting as (variational) encoder, decoder, and encoder
respectively. To train such a network one must connect the inputs and outputs
appropriately to compute the cost function $H$ (Eq.~\ref{eq:cost3}) and
back-propagate cost function derivatives to update the network parameters. 

%
%
Training is performed via a series of steps illustrated schematically in
Fig.~\ref{fig:network_config}. A batch of data composed of pairs of
time-series $y$ and their corresponding true \ac{GW} signal parameters $x$ are
passed as input and the following steps are applied to each element of the
batch.
\begin{enumerate}
\item The encoder $q_{\phi}$ takes both the time-series
$y$ and the true parameters $x$ defining the \ac{GW} signal. It then encodes
these instances into parameters $\mu_{q}$ defining an uncorrelated (diagonal covariance matrix)
$n_z$-dimensional Gaussian distribution in the latent space. 
\item The encoder $r_{\theta_1}$ is given only the time-series data
$y$ and encodes it into a set of variables $\mu_{r_1}$ defining a
multi-component multivariate Gaussian mixture distribution
in the latent space.
\item We then draw a sample from the distribution described by $\mu_{q}$ giving us
a location $z_{q}$ within the latent space.
\item This sample, along with its corresponding $y$ data, are then passed
as input to the decoder $r_{\theta_2}$. This decoder outputs $\mu_{\theta_2}$
comprising a set of parameters that define a distribution in the physical $x$
space. 
\item The first term of the loss function, the reconstruction loss (defined as $L$ in
Eq.~\ref{eq:cost3}), is then computed by evaluating the probability density
defined by $\mu_{\theta_2}$ at the true $x$ training value (the average
is then taken over the batch of input data). 
\item The second loss component, the \ac{KL}-divergence between
the distributions $q_{\phi}(z|x,y)$ and $r_{\theta_1}(z|y)$ (described by
the parameter sets $\mu_{q}$ and $\mu_{r_1}$), is approximated as 
\begin{align}\label{eq:klgauss}
\text{KL}&\left[ q_{\phi}(z|x_{n},y_{n})||r_{\theta_{1}}(z|y_{n})\right] \\
&\approx \left.\log\left(\frac{q_{\phi}(z|x_n,y_n)}{r_{\theta_1}(z|y_n)}\right)\right|_{z\sim
q_{\phi}(z|x_n,y_n)}\nonumber
\end{align}
where $z$ is the sample drawn from $q_{\phi}(z|x_n,y_n)$ in the first
training stage. We use this single-sample Monte-Carlo integration approximation
since the \ac{KL}-divergence between a single-component and a multi-component
multivariate Gaussian distribution has no analytic solution (the average
is then taken over the batch of input data). 
\item The 2 loss components are then summed according to Eq.~\ref{eq:cost3} and
all trainable network parameters (defined by $\theta_1,\theta_2,\phi$) are
updated based on the derivative of the cost function with respect to these
parameters.
\end{enumerate}

%
%
A problematic aspect of training relates to the behaviour of the network
during the initial stages of training. The network has a strong tendency to
become trapped in local minima resulting in a decreasing cost component $L$
(the reconstruction cost) but a non-evolving \ac{KL}-divergence term that
remains close to zero. To avoid this state we apply an annealing process in
which the \ac{KL}-divergence term is initially ignored but its contribution is
then increased logarithmically from 0 to 1 between the epoch indices
$1\times10^2$---$3\times10^2$. 
 This allows the $q_{\phi}$ encoder to learn the latent space
representation of the data via the reconstruction cost before being required to
try to best match its distribution to that modelled by the $r_{\theta_1}$
encoder. In parallel with the gradual introduction of the \ac{KL} cost term, we
also find that the stability of training is negatively affected by the
complexity of our tailored output decoder likelihood functions. To 
resolve this we apply the same annealing procedure over the same epoch 
range in transitioning between unbound Gaussian likelihoods on all 
physical parameters to the tailored likelihoods, where the boundaries 
of the Gaussian likelihoods are brought in from $-10$ to $0$ on the 
lower bound and $11$ to $1$ on the upper bound.

%
%
As is standard practice in machine learning applications, the cost is computed
over a batch of training samples and repeated for a pre-defined number of
epochs. An epoch is defined as the point at which the network has been trained 
on a number of samples equivalent to $2\times10^4$. 
For our purposes, we found that $\sim 3 \times 10^4$ training
epochs, a batch size of $1500$ training samples and a learning rate of
$10^{-4}$ was sufficient. We used a total of $10^7$ training samples in order
to adequately cover the \ac{BBH} parameter space. We additionally ensure that
an (effectively) infinite number of noise realizations are employed by making
sure that every time a training sample is used it is given a unique noise
realisation despite only having a finite number of waveforms. Every 4 epochs we 
load a new set of $2\times10^4$ training samples. When loading in a new set we 
augment the data on the amplitude, phase and time 
of arrival by shifting all the parameters randomly within the bounds defined by the prior.

%
%
Completion of training is determined by comparing output posteriors on test
samples with those of \texttt{Bilby} iteratively during training. This
comparison is done using standard figures of merit such as the \ac{PP}-plot
\ac{JS}-divergence (see Figs.~\ref{fig:pp_plot} and
\ref{fig:kl_results}). We also assess training completion based on whether the
evolution of the cost function and its component parts
(Fig.~\ref{fig:loss_log}) have converged. We use a single Nvidia Tesla V100
\acp{GPU} with $16/32$ Gb of RAM although consumer grade ``gaming" \ac{GPU}
cards are equally fast for this application.

\subsection{The testing procedure}
%
%
After training has completed and we wish to use the network for inference we
follow the procedure described in the right hand panel of
Fig.~\ref{fig:network_config}. Given a new $y$ data sample (not taken from the
training set) we simply input this into the trained encoder $r_{\theta_1}$ from
which we obtain a single value of $\mu_{r_1}$ describing a distribution
(conditional on the data $y$) in the latent space. We then repeat the following
steps:

%
%
\begin{enumerate}
\item We randomly draw a latent space sample $z_{r_1}$ from the latent space
distribution defined by $\mu_{r_1}$.
\item The $z_{r_1}$ sample and the corresponding original $y$ data are fed as input to our
pre-trained decoder network $r_{\theta_2}$. The decoder network returns a set
of parameters $\mu_{r_2}$ which describe a multivariate distribution in the physical
parameter space.
\item We then draw a random $x$ realisation from that distribution.
\end{enumerate}
%

%
%
A comprehensive representation in the form of samples drawn from the entire
joint posterior distribution can then be obtained by simply repeating this
procedure and hence sampling from our latent model $r_{\theta}(x|y)$ (see
Eq.~\ref{eq:latent_model}).

\subsection{Additional tests}
%
%
\begin{figure}
    \includegraphics[width=\columnwidth]{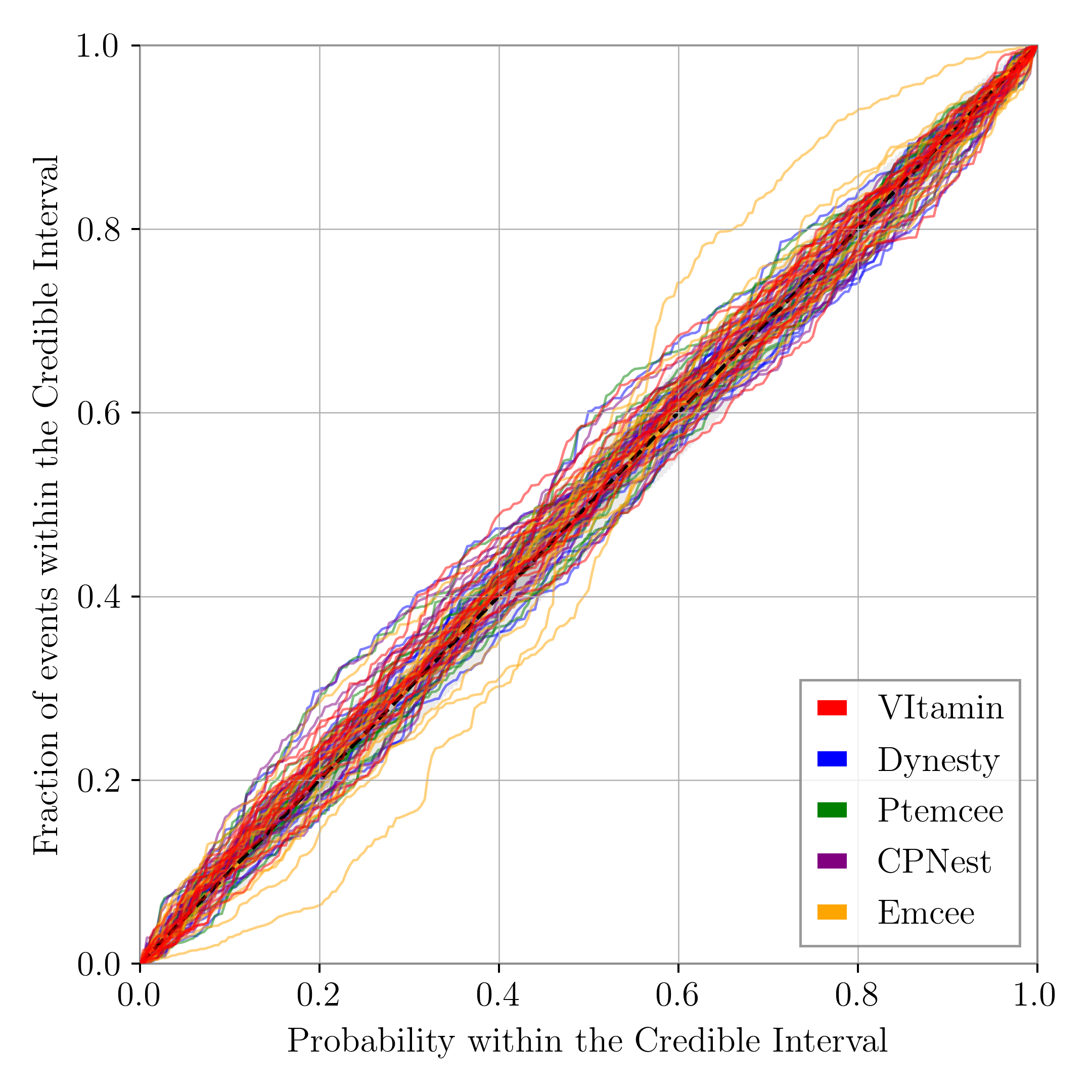}
    \caption{\label{fig:pp_plot} One-dimensional \ac{PP} plots for each
parameter and for each benchmark sampler and \texttt{VItamin}. The curves were
constructed using the 250 test datasets and the dashed black diagonal line
indicates the ideal result. The best and worst-case $p$-values associated with each
sampling method are (0.918,  0.047 \texttt{VItamin}), (0.912, 0.007 \texttt{Dynesty}), (0.931,0.007
\texttt{ptemcee}), (0.706,0.007 \texttt{CPNest}), (0.667,0.004 \texttt{emcee} ). 
}
\end{figure}
%

%
%
\begin{table*}
\centering
\caption{Benchmark sampler configuration parameters. Values were chosen based
on a combination of their recommended default parameters~\cite{1811.02042} and
private communication with the \texttt{Bilby} development team.}
\begin{tabular}[t]{lc}
\toprule
sampler & parameters \\
\hline
\texttt{Dynesty}~\cite{dynesty} & live-points $=1000$, dlogz $=0.1$, nact $=50$,  npool $=8$, bound $=$ None, sample $=$ uniform\\

\texttt{ptemcee}~\cite{ptemcee} & $\begin{array}{c}\text{walkers}=200,\, 
\text{temperatures}=20,\, \text{burn}\_\text{in}\_\text{nact}=50,\, \text{thin}\_\text{by}\_\text{nact}=0.5,\, \\
\text{nsamples}=10000,\, \text{threads}=10,\, \text{autocorr}\_\text{tol}=50,\, \text{autocorr}\_\text{c} \text{safety}=1,\, \text{autocorr}\_\text{tau}=1,\, \\ 
\text{gradient}\_\text{tau}=0.1,\, 
\text{gradient}\_\text{mean}\_\text{log}\_\text{posterior}=0.1,\, \text{Q}\_\text{tol}=1.01,\, 
\text{min}\_\text{tau}=1,\, \text{threads}=1,\, \end{array}$ \\

\texttt{CPNest}~\cite{cpnest} & live-points $=2048$, maxmcmc $=1000$, nthreads$=$ 1, 
seed$=1994$, dlogz $=0.1$ \\

\texttt{emcee}~\cite{emcee} & nwalkers $=250$, nsteps $=14000$, nburn$=4000$, a$=1.4$,
burn$\_$in$\_$fraction$=0.25$, burn$\_$in$\_$act$\_$=3 \\
\botrule
\end{tabular}
\label{Tab:sampler_params}
\end{table*}

%
%
A standard test used within the \ac{GW} parameter estimation community is the
production of so-called \ac{PP} plots which we show for our analysis and
the benchmark comparisons in Fig.~\ref{fig:pp_plot}. The plot is constructed
by computing a cumulative probability for each 1-dimensional marginalised
test posterior evaluated at the true simulation parameter value (the fraction
of posterior samples $\leq$ the simulation value). We then plot the
cumulative distribution of these values~\cite{1409.7215}. Curves consistent
with the black dashed diagonal line indicate that the 1-dimensional Bayesian
probability distributions are consistent with the frequentist interpretation -
that the truth will lie within an interval containing $X\%$ of the posterior
probability with a frequency of $X\%$ of the time. It is clear to see that
results obtained using \texttt{VItamin} show deviations from the diagonal
that are entirely consistent with those observed in all benchmark samplers.
The $p$-value has also been calculated for each sampler and each parameter
under the null-hypothesis that they are consistent with the diagonal. These
results show that for at least 1 parameter, emcee shows inconsistency with the
modal at the 0.4\% level. \texttt{Dynesty} has a worst case that is consistent only at the
0.7\% level.  All other samplers (including \texttt{VItamin}) show consistency at
$>0.4\%$ in the worst case.     

%
%
%

%
\begin{figure*}
    \includegraphics[width=\textwidth]{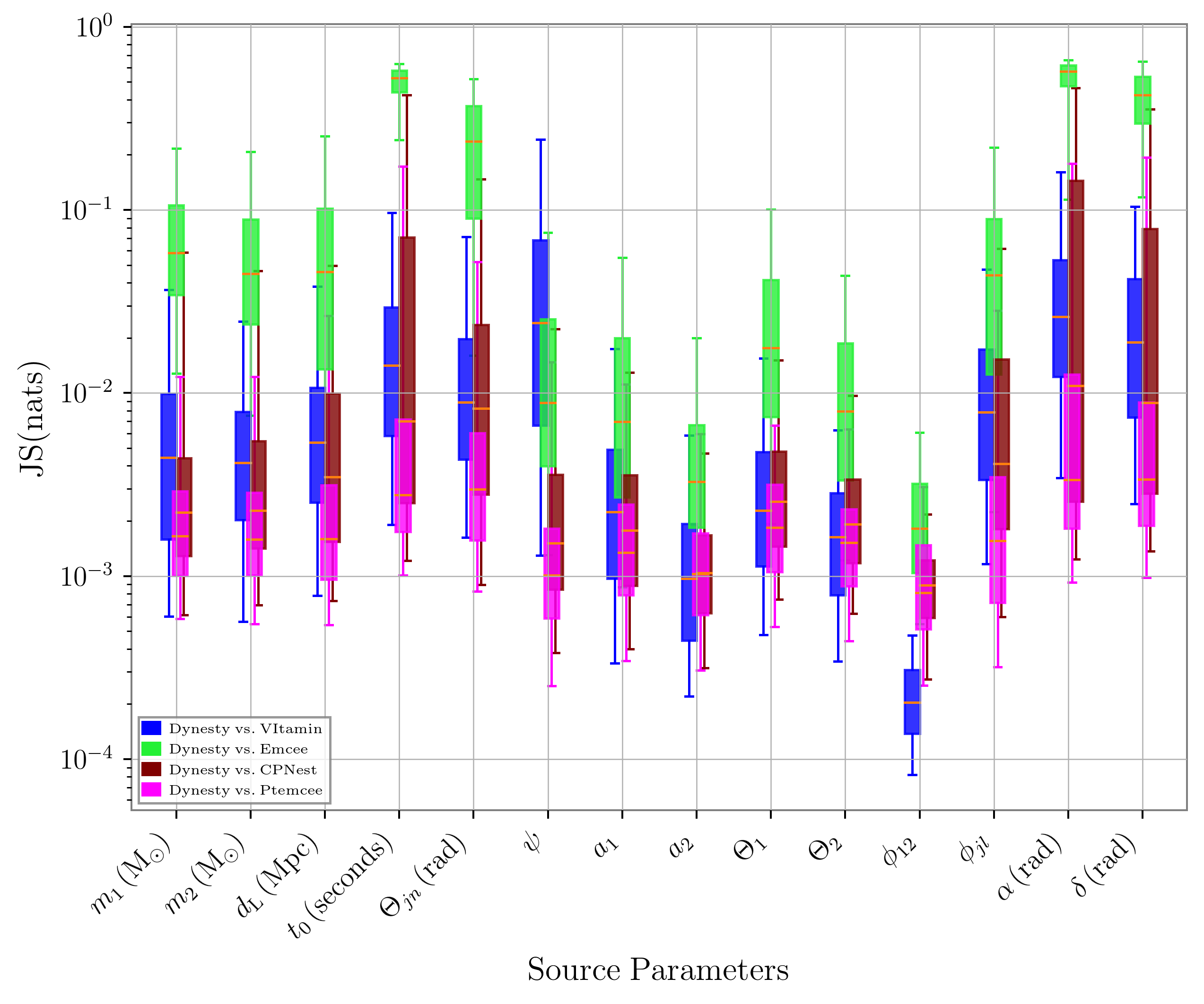}
    \caption{\label{fig:kl_results} We show JS divergence values for all 250 test samples as a function of test sample source parameter for \texttt{Dynesty} against every other sampling approach. Each sampler method vs. another sampler method are denoted as different colors. The lower and upper end of boxes represent the 25th and 75th percentile credible regions respectively. The lower and upper end of the whiskers represent the 5th and 95th percentile credible regions.  The orange lines are representative of the median JS values for each pair of compared samplers.
      }

\end{figure*}
%

%
%
The \ac{JS}-divergence is generally used as measure of the similarity 
between distributions.  In Fig.~\ref{fig:kl_results} 
we use this quantity to compare the output posterior estimates between 
samplers for the same input test data. To do this we run each independent 
sampler (including \texttt{VItamin}) on the same test data to produce 
samples from the corresponding posterior. We then compute the 
1-dimensional \ac{JS}-divergence between the output single 
parameter distributions from each sampler with every 
other sampler~\cite{4839047}. For distributions that are 
identical, the \ac{JS}-divergence \emph{should} equal zero but since we 
are representing our posterior distributions using finite numbers of 
samples, identical distributions result in \ac{JS}-divergence with finite values.  
In Fig. \ref{fig:kl_results}, it can be seen that \texttt{Dynesty}  
vs. \texttt{VItamin} JS values are competitive with results from 
\texttt{Dynesty} vs. \texttt{ptemcee} for nearly all parameters, with 
the exception of $\phi_{12}$ and $\psi$. \texttt{VItamin} predictions have slightly 
higher \ac{JS} values across all source parameters except for the 
spin parameters. The \texttt{Dynesty} vs. \texttt{CPNest} comparison seems to generally 
have similar \ac{JS} values to \texttt{Dynesty} vs. \texttt{ptemcee} with 
the exception of having broader credible intervals on $t_0$, 
$\Theta_{jn}$, $\phi_{jl}$, $\alpha$ and $\delta$. 
\texttt{Dynesty} vs. \texttt{emcee} generally has higher JS values than 
all other methods, which is expected given the difficulty of 
\texttt{emcee} convergence.

\bibliographystyle{apsrev4-1}
\bibliography{VItamin}

\end{document}